# KU-BAND ALSCN-ON-DIAMOND SAW RESONATORS WITH PHASE VELOCITY ABOVE 8600 M/S

*Tzu-Hsuan Hsu[1], Kapil Saha[2], Jack Kramer[1], Omar Barrera[1], Pietro Simoni[2], Matteo Rinaldi[2], and Ruochen Lu[1]*
[1]The University of Texas at Austin, USA, and [2]Northeastern University, USA

## ABSTRACT

In this work, an Aluminum Scandium Nitride (AlScN) on Diamond Sezawa mode surface acoustic wave (SAW) platform for RF filtering at Ku-band (12-18 GHz) is demonstrated. Thanks to the high acoustic velocity and low-loss diamond substrate, the prototype resonator at 12.9 GHz achieves a high phase velocity ($v_p$) of 8671 m/s, a maximum Bode-$Q$ of 408, and coupling coefficient ($k_{eff}^2$) of 2.1%, outperforming high-velocity substrates such as SiC and sapphire by more than 20% in velocity. Resonators spanning 8 to 18 GHz are presented. The platform's high power handling above 12.5 dBm is also experimentally validated.

## INTRODUCTION

With the growing demand for higher data rates and improved spectral efficiency in modern wireless systems, there is increasing interest in utilizing frequency bands above 10 GHz. Among them, the Ku-band (12–18 GHz) within upper-mid bands (or commonly referred as the 6G frequency range 3, i.e., FR3 bands) stands out as a promising and pivotal part of spectrum targeting the emerging next-generation wireless communication systems [1]. However, realizing efficient acoustic resonators at these high frequencies remains a significant challenge due to inherent limitations in material properties, mode confinement, and fabrication scalability [2].

Conventional surface acoustic wave (SAW) resonators based on Rayleigh modes, while well-established, suffer from limited phase velocity and poor scalability at higher frequencies [3]. To address these constraints, it is essential to explore alternative acoustic modes and employ substrates with higher acoustic velocities.

This work introduces a Sezawa mode-based AlScN-on-diamond SAW resonator platform [Fig. 1(a)-(b)] specifically engineered for Ku-band applications. As depicted in (Fig. 1(c)), AlScN is generally favorable in high frequency acoustic resonator platforms due to its high phase velocity, as seen in some recently publish millimeter wave acoustic devices [4]. However, for solidly mounted acoustic platform, the high velocity of AlScN typically indicates that only Rayleigh mode can be efficiently confined through a hetero acoustic layered setup. Common substrates such as silicon, quartz or sapphire which are widely seen in thin film lithium niobate (LN) [5, 6] or lithium tantalate (LT) [7, 8] solidly mounted acoustic platform perform poorly due to its leaky acoustic confinement caused by closed proximity of phase velocity between substrate and piezoelectric thin film, which leads to high velocity mode such as Sezawa SAW often exceeds the phase velocity of carrier substrates. Recent advancement in piezoelectric on insulator (POI) SAW has enabled silicon carbide (SiC) to be a favorable hetero acoustic substate material [9, 10] to confine Sezawa mode in thin film AlScN. As listed in Fig. 1(c) [11-13], despite promising, the mechanical property, thermal conduction and acoustic velocity of SiC are still not comparable with that promised in diamond. However, single-crystal diamond substrates have not been explored well for acoustic applications due to their limited availability and rough surfaces.

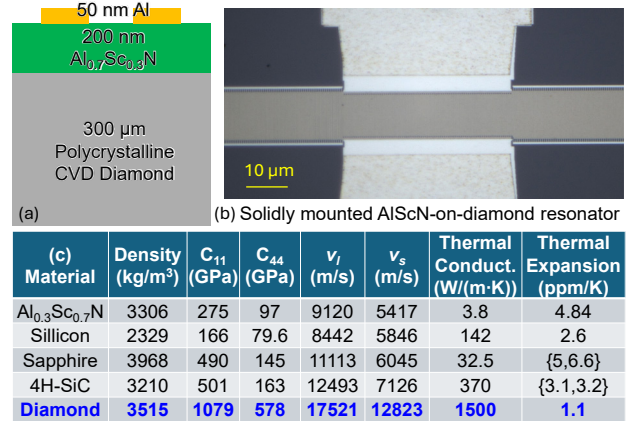

| (c) Material | Density (kg/m³) | $C_{11}$ (GPa) | $C_{44}$ (GPa) | $v_l$ (m/s) | $v_s$ (m/s) | Thermal Conduct. (W/(m·K)) | Thermal Expansion (ppm/K) |
|---|---|---|---|---|---|---|---|
| Al₀.₃Sc₀.₇N | 3306 | 275 | 97 | 9120 | 5417 | 3.8 | 4.84 |
| Silicon | 2329 | 166 | 79.6 | 8442 | 5846 | 142 | 2.6 |
| Sapphire | 3968 | 490 | 145 | 11113 | 6045 | 32.5 | {5,6.6} |
| 4H-SiC | 3210 | 501 | 163 | 12493 | 7126 | 370 | {3.1,3.2} |
| **Diamond** | **3515** | **1079** | **578** | **17521** | **12823** | **1500** | **1.1** |

*Figure 1: (a) Unit cell model of the AlScN-on-Diamond resonator, (b) optical microscope image of the fabricated SAW resonator, and (c) material property comparison between AlScN and high velocity substrates forming the acoustic waveguide.*

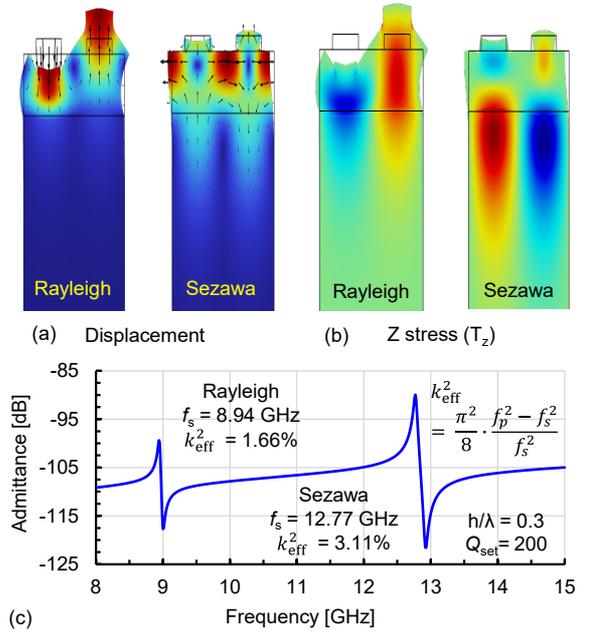

*Figure 2: (a)(b) Mode shape and stress profile of Rayleigh and Sezawa modes studied in this work, and (c) simulated Ku band resonator admittance spectrum based on a h/λ of 0.3.*

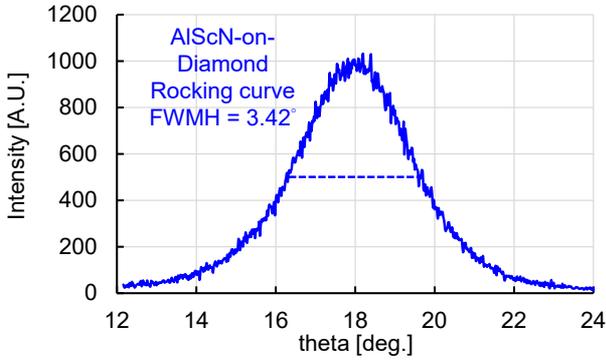

*Figure 3: X-Ray Diffraction (XRD) rocking curve measurement of the deposited AlScN thin film.*

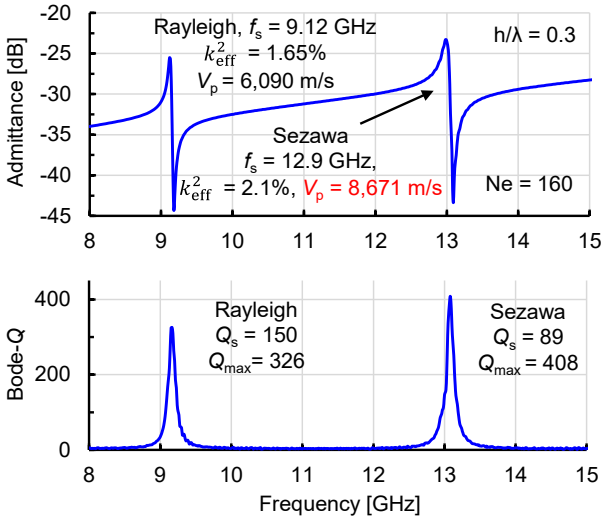

*Figure 4: Measured admittance and extracted Bode-Q for the Ku band AlScN-on-Diamond resonator.*

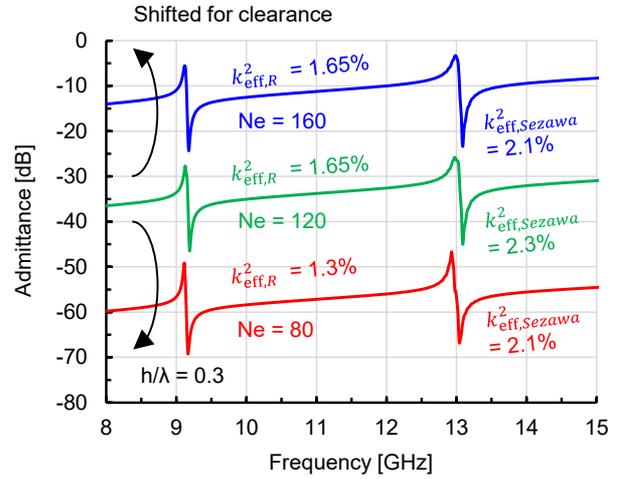

*Figure 5: Measured admittance for AlScN-on-Diamond resonator with different IDT designs.*

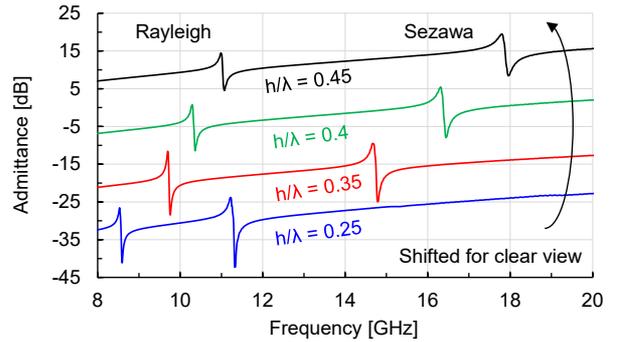

| Rayleigh | | | | Sezawa | | | |
|---|---|---|---|---|---|---|---|
| $h_{AlScN}/\lambda$ | $f_s$ [GHz] | $k_{eff}^2$ [%] | $Q_{p-3dB}$ | $h_{AlScN}/\lambda$ | $f_s$ [GHz] | $k_{eff}^2$ [%] | $Q_{p-3dB}$ |
| 0.25 | 8.5 | 1.8 | 235 | 0.25 | 11.2 | 2.15 | 310 |
| 0.35 | 9.7 | 1.6 | 200 | 0.35 | 14.7 | 2.06 | 202 |
| 0.4 | 10.3 | 1.76 | 142 | 0.4 | 16.3 | 2.04 | 150 |
| 0.45 | 11 | 1.65 | 91 | 0.45 | 16.3 | 2.04 | 150 |

*Figure 6: Measured performance for AlScN-on-Diamond resonators between h/λ of 0.25 and 0.45 (Ne = 80 for h/λ = 0.35, Ne = 160 for the rest of the resonators.)*

Recent advancement in polycrystalline diamond deposited through chemical vapor deposition (CVD) [14] has shown great potential for its much affordable price, similar mechanical characteristics to single crystalline diamond, and decent thermal conductivity (>1500W/mK), which enables diamond substrate to serve as an excellent heterogeneous acoustic waveguide platform. Therefore, our design leverages the ultra-fast, low-loss, and thermally conductive properties of diamond to effectively confine high-velocity acoustic mode within the AlScN layer without the need for dense metal electrode loading to enhance phase velocity contrast between AlScN and substrate [15, 16] and promote better confinement conditions. However, heavy electrode loading typically imposes limitations on frequency scalability, resulting in even finer electrode pitches that may feature worse power handling. On the contrary, the AlScN-on-diamond platform provides a highly scalable solution capable of supporting low-loss, high-frequency operation across the entire Ku band.

## DESIGN AND SIMULATION

To enable high-frequency operation in the Ku-band, this work adopts a Sezawa mode SAW resonator platform utilizing a 200 nm thick sputtered $Al_{0.7}Sc_{0.3}N$ thin film on a 300 μm polycrystalline diamond substrate. 50 nm Aluminum electrodes are selected as they apply minimal phase velocity loading on the targeted Sezawa mode. The heteroacoustic waveguide formation using diamond substrate ensures proper acoustic confinement without resorting to thick or heavy electrode loading.

Finite element simulations were conducted to analyze modal behavior and validate the design approach. Fig. 2(a)-(b) shows the simulated mode shape displacement profiles and the out-of-plane (Z-direction) stress profiles of both Rayleigh and Sezawa modes. The Sezawa mode clearly showed deeper penetration towards the substrate and therefore explained the significant difficulty in effectively confining Sezawa mode within solidly mounted waveguides.

Fig. 2(c) depicted the unit cell simulation result for a resonator with normalized AlScN thickness to wavelength ratio (h/λ) of 0.3. Unit cell simulation setup follows detailed procedure listed in [17] with a quality factor ($Q_{set}$) of 200. The admittance spectrum reveals Rayleigh resonance at 8.94 GHz and Sezawa resonance at 12.77 GHz. The electromechanical coupling coefficients ($k_{eff}^2$) that are then calculated through (1) are 1.66% and 3.11% respectively [18].

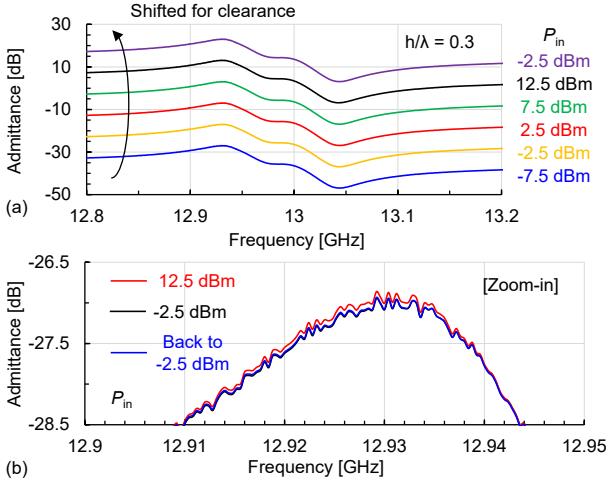

*Figure 7: Measured (a) power handling for AlScN-on-Diamond resonator with h/λ of 0.3 from -2.5 dbm up to 12.5 dbm back to -2.5 dbm (Cable loss already calibrated) and TCF, and (b) Zoom-in measurement result at the series resonance peak comparing nominal (-2.5dBm) and peak input power (12.5 dBm).*

$$k_{eff}^2 = \frac{\pi^2}{8}\frac{C_m}{C_o} = \frac{\pi^2}{8}\frac{f_p^2 - f_s^2}{f_s^2} \quad (1)$$

The simulation results confirm the feasibility of achieving Ku-band frequency upscale using AlScN-on-diamond platform and highlight the platform's potential for wide frequency tunability.

## MEASUREMENT RESULTS

The fabricated AlScN-on-Diamond resonator is shown in Fig. 1(b), consisting of 160 electrode pairs (Ne), 120 pairs of reflective gratings, and an aperture of 30λ. Due to the extraordinarily hard nature of diamond, the difficulty in surface machining of diamond is widely known. However, poor substrate surface roughness can lead to unsatisfactory deposited AlScN crystal quality. In this work, the polycrystalline diamond substrate features a surface roughness (Ra) less than 20 nm. The AlScN quality were first tested and the rocking curve (RC) of the wurtzite crystal structure within the sputtered Al₀.₇Sc₀.₃N thin film was obtained through X-ray diffraction (XRD) measurement and the result shown in Fig. 3 reveals a good crystallinity RC angle of 3.42°.

The fabricated device was then measured using a Keysight P5028A Vector Network Analyzer (VNA) with MPI TS150 GSG prober under ambient conditions. Only standard SOLT calibration was performed prior to each measurement. Frequency response measurement for a resonator with an h/λ ratio of 0.3 is depicted in Fig. 4 and the results reveal Rayleigh mode at 9.12 GHz and Sezawa mode at 12.9 GHz, respectively. The Sezawa mode achieves an extraordinary phase velocity over 8600 m/s, thus enabling excellent scalability to Ku band and above. Extracted $k_{eff}^2$ are 1.65% and 2.1% for Rayleigh and

*Table 1: Comparison with state-of-the-art.*

| Ref. | Platform | Mode | Substate Material | $f_s$ [GHz] | $V_p$ [m/s] | $k_{eff}^2$ [%] | $Q_{max}$ |
|---|---|---|---|---|---|---|---|
| JMEMS '24 | AlScN SAW | Sezawa | SiC | 5.9 | 5660 | 4 | 762 |
| IEEE EDL '24 | AlScN SAW | Sezawa | SiC | 16.2 | 6490 | 4 | 380 |
| Jpn. J. Appl. Phys. '23 | AlScN SAW | Sezawa | Diamond | 3.73 | 7460 | 5.4 | 311 |
| IEEE IC MAM '24 | LN SAW | SH-SAW | Si | 13.4 | 3210 | 7 | 58 |
| This work | AlScN SAW | Sezawa | Diamond | 12.9 | 8671 | 2.1 | 408 |

Sezawa modes, respectively. The bode-Q ($Q_{max}$) extracted using (2) is also shown in Fig. 4 with the Sezawa mode resonator achieved a high $Q_{max}$ of 408 [19, 20].

$$Q_{Bode} = \omega \cdot \left|\frac{dS_{11}}{d\omega}\right| \cdot \frac{1}{1 - |S_{11}|^2} \quad (2)$$

Variously different interdigital transducer (IDT) designs with Ne of 80, 160 and 240 under h/λ = 0.3 were also experimentally validated. The results depicted in Fig. 5 show a $k_{eff}^2$ saturating at around 1.65% for Rayleigh mode with a $k_{eff}^2$ capping at 2.3% for the Sezawa mode.

The frequency scalability potential for AlScN-on-diamond platform is then demonstrated in Fig. 6 where a series of resonators with h/λ ranging between 0.25 to 0.45 were fabricated and experimentally tested. The two distinct resonances of Rayleigh and Sezawa mode have successfully achieved frequencies spanning the entire X-band (8-12 GHz) and Ku band (12-18 GHz) respectively. The result highlights the potential of scaling AlScN-on-diamond devices towards Ku-band and beyond. Detailed performances were documented in a sub-chart listed in Fig. 6. Unfortunately, performance degradation at higher h/λ region is still observed and it can be attributed to electrode loading at extremely small wavelength. This hereby emphasizes the importance of utilizing platforms such as AlScN-on-diamond to harness ultimate phase velocity to avoid aggressive wavelength shrinking during frequency scaling in SAW device design.

Lastly, owing to the exceptional thermal conductivity of diamond, preliminary power handling of the devices was also measured. The power handling results depicted in Fig. 7 (a) demonstrate robust power-handling capabilities. The device with h/λ=0.3 was repeatedly tested with incident power ($P_{in}$) at the device ranging between -7.5 dBm and 12.5 dBm before going back to -2.5 dBm. Cable loss was removed during calibration. As shown in Fig. 7(b), the zoom-in admittance spectra near the series resonance shows no significant performance degradation.

## CONCLUSION

To benchmark the proposed AlScN-on-diamond devices against the current state-of-the-art in solidly mounted SAW platforms, a comparison is summarized in Table 1 [9, 15, 21, 22], highlighting the advantages of the AlScN-on-diamond architecture over prior works. This platform exhibits an unparalleled high phase velocity that surpasses other high-velocity SAW substrates with comparable feature sizes. For example, achieving a similar operational frequency on a SiC-based platform would require further miniaturization up to a 20% reduction in feature size as depicted in [9]. This unique characteristic of AlScN-on-diamond underscores the platform's potential for frequency up-scaling and advanced signal processing at Ku-band and beyond.

The results demonstrate the feasibility and performance advantages of employing diamond as the substrate for AlScN-based Sezawa mode SAW resonators. The exceptionally high phase velocity of 8671 m/s, coupled with a competitive $Q_{\max}$ and $k_{\text{eff}}^2$, positions this platform as a strong candidate for next-generation high-frequency RF filters. While the $k_{\text{eff}}^2$ currently presented is moderate, it is effectively compensated by the broad frequency tunability enabled through variations in the normalized thickness h/λ.

Additionally, the demonstrated power handling results reflect promising potential in power endurance capability of the AlScN-on-diamond platform under experimental conditions. The AlScN-on-diamond platform not only offers superior acoustic velocity but also leverages diamond's exceptional thermal conductivity to alleviate self-heating that often impairs device reliability and performance. These combined advantages make this platform particularly attractive for emerging thermally demanding wireless applications like 6G FR3 and others where both performance and miniaturization are critical.

## ACKNOWLEDGEMENTS

The authors would like to acknowledge DARPA COFFEE program and NSF CAREER 2339731: Radio Frequency Piezoelectric Acoustic Microsystems for Efficient and Adaptive Front-End Signal Processing for funding support. The authors would like to thank Dr. Benjamin Griffin, Dr. Todd Bauer, and Dr. Zachary Fishman for helpful discussions.

## CONTACT

*Tzu-Hsuan Hsu; tzuhsuan.hsu@austin.utexas.edu